\RequirePackage{latexml}
\iflatexml
\documentclass{article}
\usepackage{amsmath,amssymb,graphicx,booktabs,multirow,longtable,braket,array,tabularx}
\usepackage[numbers,sort&compress]{natbib}
\usepackage{hyperref,xcolor,authblk}
\definecolor{QKPrimary}{HTML}{19364A}
\title{Fully GPU-accelerated semistochastic Epstein--Nesbet perturbation theory}
\author[1]{Lizhong Fu}
\author[1]{Honghui Shang\thanks{Corresponding author: \href{mailto:shanghui.ustc@gmail.com}{shanghui.ustc@gmail.com}.}}
\author[1]{Jinlong Yang\thanks{Corresponding author: \href{mailto:jlyang@ustc.edu.cn}{jlyang@ustc.edu.cn}.}}
\affil[1]{Key Laboratory of Precision and Intelligent Chemistry, University of Science and Technology of China, Hefei, Anhui 230026, China}
\date{10 November 2026}
\newenvironment{qkfrontmatter}{}{}
\newcommand{\qkbibliography}[1]{
}
\newcommand{\qkappendix}{\appendix}
\newcommand{\FloatBarrier}{}
\newcommand{\Needspace}[1]{}
\newcommand{\balance}{}
\newcommand{\nobalance}{}
\providecommand{\pdfbookmark}[3][]{}
\makeatletter
\let\ps@qkbody\ps@plain
\makeatother

\else
\documentclass[paper]{qiankunnet}
\usepackage{braket,longtable}
\AtBeginEnvironment{thebibliography}{\nobalance}
\makeatletter
\patchcmd{\@lbibitem}{\bibitem@fin}{\bibitem@fin\ifstrequal{#2}{SafariFCIQMCCASPT22025}{\newpage}{}}{}{\ClassError{arxiv}{Bibliography layout patch failed}{}}
\makeatother

\title{Fully GPU-accelerated semistochastic Epstein--Nesbet perturbation theory}
\qkpdftitle{Fully GPU-accelerated semistochastic Epstein--Nesbet perturbation theory}
\author{Lizhong Fu\quad Honghui Shang\textsuperscript{*}\quad Jinlong Yang\textsuperscript{*}}
\qkaffiliations{Key Laboratory of Precision and Intelligent Chemistry,
University of Science and Technology of China, Hefei, Anhui 230026, China}
\qkauthornote{* Corresponding authors:
\href{mailto:shanghui.ustc@gmail.com}{shanghui.ustc@gmail.com} (Honghui Shang);
\href{mailto:jlyang@ustc.edu.cn}{jlyang@ustc.edu.cn} (Jinlong Yang).}
\qkpdfauthor{Lizhong Fu; Honghui Shang; Jinlong Yang}
\date{10 November 2026}
\fi
\newcommand{\ha}{E_{\mathrm h}}
\newcommand{\uha}{\mu E_{\mathrm h}}
\newcommand{\mha}{\mathrm{m}E_{\mathrm h}}
\newcommand{\Sref}{\mathcal S}
\newcommand{\Dref}{\mathcal D}
\newcommand{\Tref}{\mathcal T}
\newcommand{\normc}{\mathcal N}

\begin{document}
\begin{qkfrontmatter}
\maketitle
\begin{abstract}
Second-order Epstein--Nesbet perturbation theory (EN-PT2) is an important tool for improving selected configuration interaction (SCI) energies, but the large external determinant space creates substantial memory and communication costs on GPU clusters.
We develop a fully GPU-accelerated implementation of distributed semistochastic EN-PT2, fusing excitation generation, screening, and PT2 weighting in a single kernel traversal. Fixed-size random-phase CountSketch arrays limit memory use and replace large external determinant exchange with bounded collective summation.
The implementation on one A100 completes EN-PT2 calculations 8.21--37.53 times faster than Arrow on 64 CPU cores at the same requested statistical standard error, and fixed-workload strong-scaling tests achieve 93.5\% parallel efficiency on 16 GPUs across two nodes.
Applications to an iron--oxo complex in a (60e, 187o) active space, iron--porphyrin references containing 200 million determinants per state, and neural-network quantum states establish the method's efficiency for perturbative corrections to large reference wave functions in the GPU era.
\end{abstract}
\end{qkfrontmatter}

\section{Introduction}
Selected configuration interaction (SCI) methods represent the wave function as a linear combination of selected Slater determinants~\cite{cipsi,asci,TubmanModernASCI2020,hci,shci,SchriberACI2016,ZhangiCIPT22020,WilliamsYoungParallelASCI2023,DangParallelHCI2023,WangMPIiCIPT22026} and are important tools for large-scale calculations of correlated electronic systems. Second-order Epstein--Nesbet perturbation theory (EN-PT2, or PT2) is central to obtaining accurate energies with these methods. Starting from a variational wave function in a selected determinant space, the Epstein--Nesbet correction estimates the correlation energy associated with determinants outside that space~\cite{Epstein1926,Nesbet1955}. It recovers additional correlation energy without enlarging the space used for diagonalization and, together with systematic expansion of the reference, provides a route toward the full configuration interaction limit.

The PT2 calculation can be expensive. Each reference determinant connects to many other determinants through single and double excitations, and the number of distinct external determinants can greatly exceed the reference size. Contributions from different reference determinants to the same external determinant must be added before taking the squared modulus in the energy expression. Explicit generation and storage of these contributions therefore require substantial computation and memory~\cite{TubmanDeterministicPT2018,TubmanModernASCI2020,ZhangPerturbativeCDFCI2026}. Stochastic and semistochastic algorithms reduce these costs through importance sampling, screening, and a combination of deterministic and stochastic evaluation~\cite{shci,GarnironHybridPT22017,LiFastSHCI2018,GarnironQuantumPackage2019}. Arrow implements the fast semistochastic heat-bath configuration interaction (SHCI) method~\cite{hci,shci,LiFastSHCI2018,ArrowSoftware2026}.

Leading supercomputers now draw much of their computing capacity from GPUs~\cite{TOP500June2026}, and variational wave-function methods based on sparse determinant expansions are increasingly implemented on GPU clusters. Trimmed configuration interaction (TrimCI) searches for a compact determinant core and then expands it, using GPUs for large diagonalizations~\cite{TrimCICOO2026}. Learned configuration selection and neural-network-guided SCI use classifiers or generative models to propose determinants~\cite{CoeMLCI2018,CoeMLCI2019,PinedaFloresChembot2021,CasierMLSCI2026,SchmerwitzNNCI2025,ThirionNaturalOrbitalNNCI2025,ZhangHAARSCI2025,SunCuNNQSSCI2026,KanGTNNSCI2025,KanNNQSSCI2025}. Neural-network quantum state (NNQS) methods generate determinants by sampling the network, and some evaluate projected energies on the sampled determinants~\cite{Wu2023_SC,MalyshevPeakedNQS2024,BarrettLvovsky2022,ShangQiankunNet2025}. These methods therefore provide explicit determinant references to which an external Epstein--Nesbet correction can be applied~\cite{SharmaStochasticPT2018,LiCr2NQS2024,CheDeterministicNNQS2026}. An efficient GPU implementation would make this correction directly available on the same GPU cluster used to generate the reference.

Evaluating the EN-PT2 correction on a GPU cluster faces three obstacles. First is excitation generation. Existing SCI and PT2 implementations, including codes with GPU diagonalizations, can still generate coupled determinants on CPUs~\cite{ArrowSoftware2026,TrimCISoftware2026}. Efficient GPU execution is complicated by indirect excitation-table accesses and a source-dependent number of excitations surviving screening. Explicitly storing these variable-length outputs requires managing buffer space and coordinating parallel writes~\cite{SunCuNNQSSCI2026}. Second is device memory. A GPU that generates the excitations has to store the reference determinants, the integral tables, and the partial sums of the external contributions. These partial sums contain far more determinants than the reference. Third is communication across GPUs, when a distributed implementation is considered. An external determinant may be reached from reference determinants processed on different GPUs, so its contributions have to be brought together before the squared modulus is taken. Grouping explicit external records across GPUs requires communication, commonly an irregular all-to-all exchange, whose volume grows with the number of generated records. Scalable calculations therefore need efficient GPU excitation generation, an accumulator that fits in device memory, and a way to combine its contents without exchanging those records.

In this work, we develop a fully GPU-accelerated distributed semistochastic EN-PT2 implementation for large configuration interaction wave functions, with limited memory use and no exchange of external determinant records. A fused kernel generates and screens excitations directly on GPUs, and computes their denominator-weighted contributions in a single traversal. These contributions are accumulated directly in a fixed-size random-phase variant of CountSketch~\cite{AlonMatiasSzegedy1996,CharikarCountSketch2004,IitakaRandomPhase2004,TrimCISoftware2026}, avoiding storage of the full external determinant set. The same bucket and phase hashes on every GPU allow these arrays to be summed by AllReduce, without exchanging external determinant records. This implementation accepts explicit determinants and coefficients from different variational methods, and can thus be used with any SCI method on GPU clusters. We compare the PT2 results with previous CPU implementations and examine runtime speedup and parallel performance, before demonstrating its application to large active space calculations as well as NNQS reference states.

\section{Semistochastic perturbation theory}
\label{sec:method}
\subsection{EN-PT2 energy correction}
For a finite determinant reference $\Sref$, let $c_i$ be its coefficients, $\normc=\sum_{i\in\Sref}|c_i|^2$, and $E_{\Sref}$ its variational energy. The reference and Hamiltonian use the same orbitals and core-energy convention. The Epstein--Nesbet correction~\cite{Epstein1926,Nesbet1955,cipsi,hci,shci} sums over determinants outside $\Sref$. With heat-bath screening~\cite{hci,LiFastSHCI2018}, the correction is
\begin{equation}
 \begin{aligned}
 F(\epsilon)&=\frac{1}{\normc}\sum_{a\notin\Sref}
 \frac{|\sum_{i\in\Sref}t_{ai}^{\epsilon}|^2}{\widetilde d_a},\\
 t_{ai}^{\epsilon}&=H_{ai}c_i\,\mathbf1_{|H_{ai}c_i|\geq\epsilon}\mathbf1_{H_{ai}c_i\ne0},\\
 d_a&=E_{\Sref}-H_{aa}.
 \end{aligned}
 \label{eq:target}
\end{equation}
Here $\widetilde d_a$ is a sign-preserving regularization of $d_a$ near zero (Appendix~\ref{app:estimator}).

\subsection{Deterministic control and sampled correction}
Following SHCI~\cite{shci,LiFastSHCI2018}, we combine a deterministic control at a coarse screening threshold $\epsilon_c=\epsilon_2^d$ with a sampled correction to the fine threshold $\epsilon_f=\epsilon_2$, where $\epsilon_c\geq\epsilon_f$:
\begin{equation}
 \begin{gathered}
F(\epsilon_f)=F(\epsilon_c)+[F(\epsilon_f)-F(\epsilon_c)],
 \\
  \epsilon_c\geq\epsilon_f.
 \label{eq:control}
 \end{gathered}
\end{equation}
The deterministic control $C=F(\epsilon_c)$ traverses all reference determinants. To estimate $\Delta=F(\epsilon_f)-F(\epsilon_c)$, we retain the $K$ largest-$|c_i|$ determinants as a fixed set $\Dref$ and draw $m\geq2$ samples with replacement from $\Tref=\Sref\setminus\Dref$. If $n_i$ is the number of times $i$ is drawn, the sampling probabilities and source weights are
\begin{equation}
 \begin{gathered}
p_i=\frac{|c_i|}{\sum_{j\in\Tref}|c_j|}\quad(i\in\Tref),\\
 w_i=\begin{cases}1,&i\in\Dref,\\n_i/(mp_i),&i\in\Tref.\end{cases}
 \end{gathered}
\end{equation}
Squaring the weighted amplitude sum introduces a sampling bias, which an analytic self-term correction removes (Appendix~\ref{app:sampled-reference}). Each iteration uses the same draws for the two thresholds to reduce the variance of their difference.

The external space is divided into $P$ disjoint batches by a hash of each determinant's complete bit string~\cite{LiFastSHCI2018}. One batch is selected uniformly per iteration, and its correction is multiplied by $P$. Both thresholds use this same batch. Iterations continue until the combined statistical standard error (SE) of the deterministic control and sampled correction reaches the requested value.

\begin{figure*}[t]
 \centering\includegraphics[width=\linewidth]{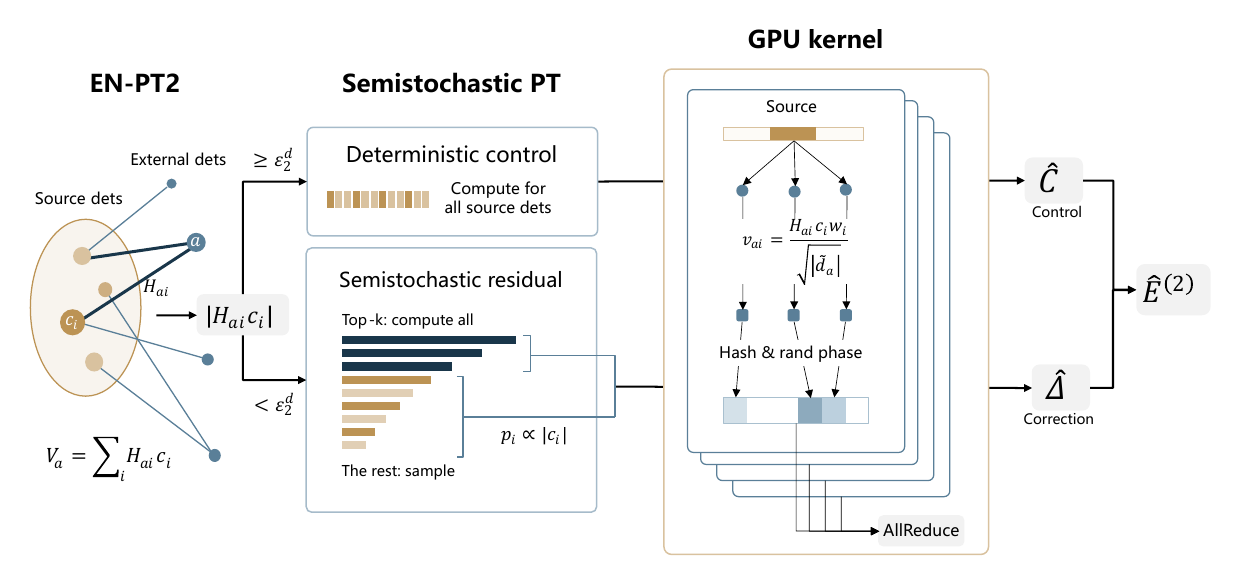}
 \caption{Overview of GPU-resident distributed semistochastic second-order Epstein--Nesbet perturbation theory (EN-PT2). The deterministic control part traverses all reference determinants to evaluate excitations with large $|H_{ai}c_i|$. The semistochastic correction part retains the $K$ reference determinants with the largest $|c_i|$ and samples the remainder. Both routes use the same GPU kernel to generate excitations, calculate their contributions, and accumulate them in sketch arrays. The arrays are summed across GPUs using AllReduce, and the EN-PT2 correction is evaluated from the combined arrays.}
 \label{fig:method}
\end{figure*}

\section{GPU implementation}
\label{sec:implementation}
The GPU implementation generates screened excitations from the reference, accumulates their contributions in fixed-size arrays, and sums those arrays across GPUs before evaluating the energy. Figure~\ref{fig:gpu-implementation} illustrates excitation generation and bucket accumulation.

\subsection{GPU excitation generation and direct accumulation in one kernel}
\label{sec:coupled-kernel}
\begin{figure*}[t]
\centering\includegraphics[width=\linewidth]{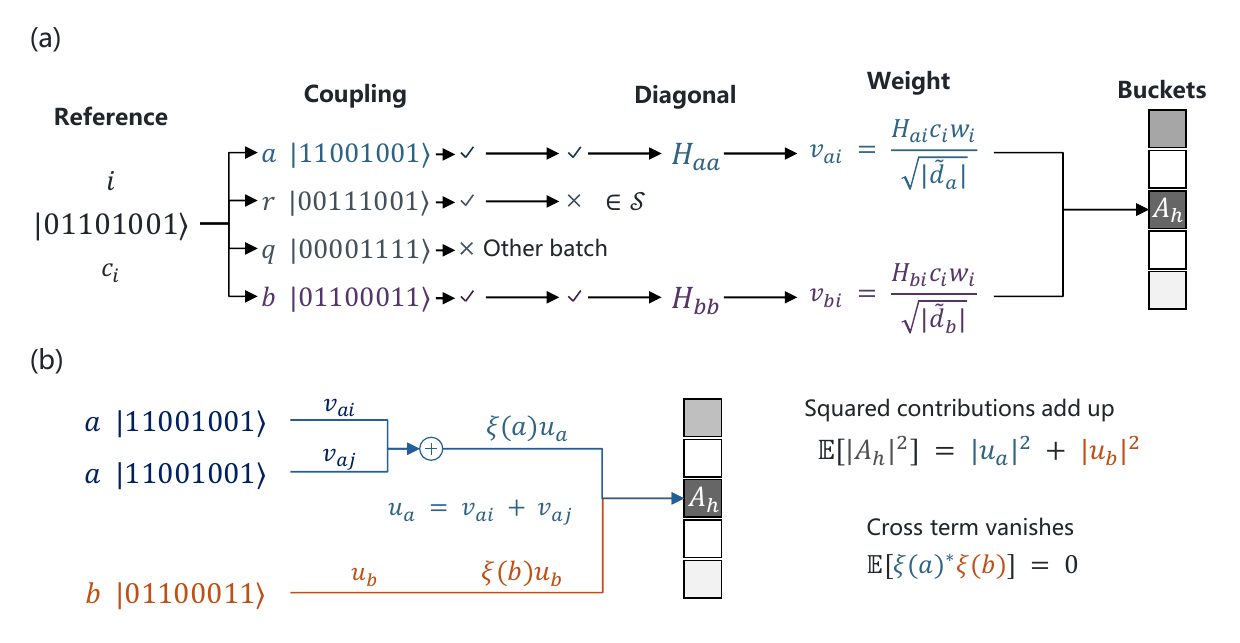}
\caption{GPU excitation generation and sketch accumulation. (a) From source determinant $i$, the kernel generates excitation configurations, rejects determinants outside the selected external batch or inside $\Sref$, evaluates each retained diagonal $H_{aa}$, and accumulates the weighted coupling into a sketch array. (b) Contributions to the same external determinant share a bucket and random phase and therefore add before squaring. Different determinants may share a bucket, but their independent phases make the cross term vanish in expectation. One screening threshold and one denominator-sign array are shown.}
\label{fig:gpu-implementation}
\label{fig:kernel-layout}
\end{figure*}

Excitation generation, screening, and direct accumulation are all implemented on the GPU in a single fused kernel. Excitation contributions are accumulated as soon as they are generated, so no explicit external determinant list is needed. For the deterministic control, each GPU processes a subset of the full reference; for the sampled correction, it processes a subset of the weighted sources defined in Section~\ref{sec:method}. The reference bit strings and coefficients, integral tables, reference-membership hash table, and accumulation arrays remain in GPU memory throughout the evaluation. Several thread blocks can scan different excitation lists for the same source determinant. Each warp handles an occupied orbital or orbital pair, with its threads scanning different table entries. Source occupations are reused in shared memory within each block. For each source determinant $i$ at screening threshold $\epsilon$, the fused kernel performs three operations (Figure~\ref{fig:gpu-implementation}a):
\begin{enumerate}
\item Generate single and double excitations from the precomputed tables, retaining only connections satisfying $|H_{ai}c_i|\geq\epsilon$ and excluding zero matrix elements and occupied destination orbitals. The double-excitation tables are sorted so that a scan stops when its entries fall below the cutoff.
\item Construct the occupation bit string of the resulting determinant $a$. Reject determinants outside the selected one of $P$ external batches or inside $\Sref$.
\item Evaluate $H_{aa}$ only for retained external determinants, apply the denominator and estimator weight, and immediately add the contribution to an accumulation array. Accumulate the sampling-bias self term in the same traversal.
\end{enumerate}

The external-batch hash test precedes the reference lookup, so discarded candidates incur no membership query, diagonal evaluation, or accumulation. Retained contributions are written directly to the fixed-size sketch arrays; no external determinant records are stored for later sorting or grouping. Section~\ref{sec:sketch} explains the sketch construction and its statistical properties.

For the sampled correction, the fused kernel evaluates both screening thresholds in a single traversal. In each sorted double-excitation list, connections satisfying $|H_{ai}c_i|\geq\epsilon_c$ update both arrays; after crossing $\epsilon_c$, only the fine-threshold array is updated until the scan stops below $\epsilon_f$. Each thread records this transition along its ordered scan, avoiding further coarse-threshold comparisons in the fine-only part. Single excitations are tested individually because their matrix elements depend on the occupied orbitals. Both thresholds reuse the matrix element, determinant bit string, reference lookup, and denominator. The deterministic control uses the same operations at $\epsilon_c$ alone.

\subsection{External contribution accumulation with CountSketch}
\label{sec:sketch}
Excitation contributions are accumulated directly in a random-phase variant of CountSketch~\cite{AlonMatiasSzegedy1996,CharikarCountSketch2004,IitakaRandomPhase2004,TrimCISoftware2026}. Arrays of $B$ sketch buckets are allocated on each GPU. Hashes of the bit string assign each $a$ a bucket $h(a)\in\{1,\ldots,B\}$ and a phase $\xi(a)\in\{1,-1,\mathrm i,-\mathrm i\}$. At one threshold, let $v_{ai}=t_{ai}^{\epsilon}w_i/\sqrt{|\widetilde d_a|}$ and $u_a=\sum_i v_{ai}$. Positive and negative denominators use separate arrays, so with $s_a=\operatorname{sgn}(\widetilde d_a)$ the stored entries are
\begin{equation}
 A_{s,b}=(\mathcal Ku)_{s,b}
 =\sum_{\substack{a\notin\Sref\\s_a=s,\;h(a)=b}}\xi(a)u_a .
 \label{eq:sketch}
\end{equation}

Contributions to the same determinant have the same hash and the same phase, and thus add together (Figure~\ref{fig:gpu-implementation}b). Different determinants that collide in the same bucket introduce cross terms, but their expectation vanishes because the phases are independent. Consequently, under the ideal random-map assumptions,
\begin{equation}
 \begin{gathered}
Q(A)=\sum_{b=1}^{B}\bigl(|A_{+,b}|^2-|A_{-,b}|^2\bigr),
 \\
  \mathbb E_{h,\xi}[Q(A)]=\sum_{a\notin\Sref}s_a|u_a|^2.
 \label{eq:sketch-expectation}
 \end{gathered}
\end{equation}
For each threshold $\ell\in\{f,c\}$, let $A_{\ell,D}$ and $A_{\ell,T}$ be the sketches formed from fixed and sampled sources in the selected external batch $\mathcal B$. The bias-corrected estimator is
\begin{equation}
 \begin{aligned}
 \widehat F_\ell&=\frac{P}{\normc}\Bigl[Q(A_{\ell,D}+A_{\ell,T})\\
 &\qquad+\frac{Q(A_{\ell,T})}{m-1}-S_\ell\Bigr],\\
 S_\ell&=\sum_{\substack{i\in\Tref\\n_i>0}}
 \left(1-\frac{mp_i}{n_i}+\frac{1}{m-1}\right)w_i^2\\
 &\qquad\sum_{a\in\mathcal B}\frac{|t_{ai}^{\epsilon_\ell}|^2}{\widetilde d_a}.
 \end{aligned}
 \label{eq:semistochastic-estimator}
\end{equation}
The self term $S_\ell$ is accumulated directly without sketching. Averaging independent control and sampled-correction evaluations gives $\widehat F=\overline C+\overline{\widehat F_f-\widehat F_c}$; the reported SCI+PT2 energy is $E_{\Sref}+\widehat F$. For fixed evaluation counts and ideal random maps, $\mathbb E[\widehat F]=F(\epsilon_f)$; screening and regularization errors are separate from the reported statistical SE. Appendix~\ref{app:sampled-reference} gives the derivation.

This implementation avoids storing the large external determinant set in full. The array size depends on $B$, rather than on the number of external determinants generated. It also avoids sending the external determinants to gather contributions from different GPUs that connect to the same external determinant. Contributions to the same determinant on each GPU enter the same bucket and have the same phase, so an NVIDIA Collective Communications Library (NCCL) AllReduce~\cite{NCCLDocumentation} sums them coherently before their squared moduli are evaluated. 

\subsection{Standard error, cost model and scheduling}
\label{sec:cost-scheduler}
Statistical uncertainty arises from sampling reference sources and external batches in the semistochastic method, and from hash collisions in the CountSketch implementation. Let $R$ denote the number of independent sketches averaged for each set of contributions, and $W$ the number of independent sampled-correction iterations. These contributions combine as
\begin{equation}
 \mathrm{SE}^2\simeq V_C+
 \frac{1}{W}\left(V_{\rm sample}+\frac{V_{\rm hash}(B)}{R}\right),
 \label{eq:variance-cost}
\end{equation}
where $V_C$ is the variance of the averaged deterministic control, $V_{\rm sample}$ is the variance over source draws and external batches after averaging over hashes, and $V_{\rm hash}(B)$ is the mean one-sketch variance at fixed sampled work. Under the ideal random-map assumptions, $V_{\rm hash}(B)=O(B^{-1})$. The reported SE combines a control variance bound with the observed variance of the sampled correction (Appendix~\ref{app:variance}).

The complete PT2 runtime is approximately
\begin{equation}
 T_{\rm PT2}\simeq T_{\rm setup}+n_c\,t_c(B_c,R)
                  +W\,t_\Delta(m,K,P,B,R),
 \label{eq:time-cost}
\end{equation}
where $T_{\rm setup}$ covers parameter selection, $n_c$ counts control evaluations, and $t_c$ and $t_\Delta$ are the mean times per control and sampled-correction evaluation.

With $B_c$ buckets per denominator sign for the deterministic control, the peak memory use is approximately
\begin{equation}
 \begin{aligned}
 M_{\rm peak}&\simeq M_0+\max\!\left(32RB_c,\right.\\
 &\qquad\left.160RB+M_{\rm sample}\right)\quad\text{bytes},
 \end{aligned}
 \label{eq:memory-cost}
\end{equation}
where $M_0$ contains the resident reference, membership index, sampling arrays, and integral tables, and $M_{\rm sample}$ is additional sampling workspace. Each amplitude field stores two complex FP64 values per bucket, one per denominator sign, and thus uses $32RB$ bytes. The sampled correction retains four fields (two thresholds, each with fixed and sampled sources), requiring $128RB$ bytes, plus one temporary field for $A_{\ell,D}+A_{\ell,T}$, giving $160RB=128RB+32RB$. A smaller requested SE may require more sampling work through $m$, $K$, or $W$, or a larger $BR$ to reduce collision noise. More iterations increase runtime without increasing peak memory, whereas larger sampling workspaces and sketch arrays also increase memory use. Comparisons between implementations should therefore use the same requested statistical SE.

For the semistochastic calculations, we fix $R=1$, $P=16$, and an expected distinct-source target $U=262144$, including $K=\lfloor U/4\rfloor$ fixed sources, and determine $m$ to reach this target. The scheduler then selects the largest power-of-two bucket counts $B_c$ and $B$ that fit the GPU memory budget after accounting for resident data, workspaces, and safety reserves. The deterministic control is reevaluated with independent hashes if its variance bound exceeds half the requested variance budget, although no repetition was needed in our tests. The sampled correction uses at least six iterations and continues until the combined reported SE reaches the target.

\section{Results}
\label{sec:results}
\subsection{Numerical validation}
Table~\ref{tab:correctness} compares the PT2-corrected energies from the GPU implementation and Arrow for 10 molecules. LiH, H$_2$O, NH$_3$, C$_2$, and CO use all-electron STO-3G spaces; N$_2$ uses frozen-core cc-pVTZ (10e58o). The Cr$_2$ and aromatic-ring Hamiltonians are also used in the performance comparison and are specified below. Each pair uses the same reference determinants and coefficients, molecular orbitals, integrals, reference energy, and final screening threshold. The reference coefficients are obtained using Arrow. The corrected energies closely agree across references ranging from tens to millions of determinants. The table lists each program's reported SE.
\begin{table*}[!tp]
\centering\small
\caption{PT2-corrected total energies and reported standard errors from identical reference inputs. Active spaces are given as (correlated electrons, spatial orbitals); $N_{\rm ref}$ is the number of reference determinants. Energy is in $\ha$ and SE is in $\uha$.}
\label{tab:correctness}
\setlength{\tabcolsep}{3pt}
\begin{tabular}{llrrrrr}
\toprule
\multirow{2}{*}{System} & \multirow{2}{*}{Active space} & \multirow{2}{*}{$N_{\rm ref}$} & \multicolumn{2}{c}{Corrected energy} & \multicolumn{2}{c}{Reported SE}\\
\cmidrule(lr){4-5}\cmidrule(lr){6-7}
 & & & This work & Arrow & This work & Arrow\\
\midrule
LiH & (4,6) & 69 & $-7.784460280031$ & $-7.784460280031$ & $0$ & $0$\\
H$_2$O & (10,7) & 128 & $-75.012520800538$ & $-75.012520800480$ & $2.20\times10^{-6}$ & $2.20\times10^{-13}$\\
NH$_3$ & (10,8) & 256 & $-55.519214759465$ & $-55.519214494753$ & $0.0153$ & $0.846$\\
C$_2$ & (12,10) & 512 & $-74.640476522219$ & $-74.640476327178$ & $0.0171$ & $0.674$\\
CO & (14,10) & 512 & $-111.363373711031$ & $-111.363372979762$ & $0.0276$ & $1.77$\\
N$_2$ & (10,58) & 391,083 & $-109.374986190745$ & $-109.374994180906$ & $0.66$ & $6.01$\\
Cr$_2$ & (28,76) & 2,000,000 & $-2099.901142459677$ & $-2099.901139824571$ & $4.1$ & $4.93$\\
Benzene & (30,108) & 2,000,000 & $-231.556532476570$ & $-231.556533051705$ & $2.98$ & $8$\\
Pyrazine & (30,98) & 2,000,000 & $-263.566097503813$ & $-263.566092953221$ & $9.46$ & $9.98$\\
Pyrimidine & (30,98) & 2,000,000 & $-263.571127482888$ & $-263.571120891852$ & $9.95$ & $9.99$\\
\bottomrule
\end{tabular}
\end{table*}

To assess the reported SE, we also performed 64 independent complete PT2 calculations for a Fe-porphyrin triplet with a one-million-determinant reference. Details of this system and the reference preparation are given in Section~\ref{sec:large-active-spaces}. The empirical standard deviation, $1.916\,\uha$, agrees with the root-mean-square reported SE, $1.924\,\uha$. At the production bucket count $B=2^{28}$, hash collisions account for approximately 2\% of the sampled-correction variance, with the remainder arising from source sampling and external-batch selection. The collision variance decreases approximately as $1/B$, with a fitted log--log slope of $-0.989$ (Appendix Figure~\ref{fig:si-variance-validation}).

\subsection{Speedup against CPU implementations}
We compare the performance using Cr$_2$, benzene, pyrazine, and pyrimidine, each with two million unique nonzero reference determinants. Cr$_2$ uses the spin-free exact two-component (X2C) Hamiltonian with cc-pVDZ-DK and frozen Ne cores, giving a 28e76o space. The three organic molecules use nonrelativistic cc-pVDZ with the C/N $1s$ cores frozen: 30e108o for benzene and 30e98o for each diazine. Benzene uses the published SHCI-orbital integrals~\cite{EriksenBenzene2020}; pyrazine and pyrimidine use RHF orbitals at published geometries~\cite{DamourRings2021}.

The GPU implementation uses one A100-SXM4-80GB and four physical CPU cores on a shared node, and Arrow uses 64 physical CPU cores on an exclusive node and an 800-GB program memory budget. Both programs request a statistical standard error of $10\,\uha$. Each program retains its original stopping rule.

\begin{figure}[tbp]
 \centering\includegraphics[width=\linewidth]{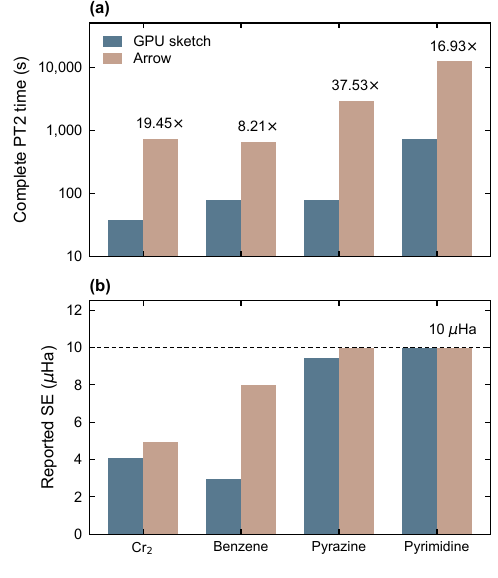}
 \caption{Runtime speedup against the CPU implementation. (a) Complete PT2 wall time for the GPU implementation on one A100 with four physical CPU cores (blue) and Arrow on 64 physical CPU cores (orange). Labels above the bars give Arrow-to-GPU speedups. (b) Reported standard errors for the same calculations; the dashed line marks the target value of $10\,\uha$.}
 \label{fig:performance}
\end{figure}

Figure~\ref{fig:performance} compares their end-to-end PT2 times, including the deterministic control and the sampled correction. The GPU implementation is faster for every system under these settings, by factors of 8.21--37.53. Because each program has a built-in minimum number of sampling iterations, some of these results overshoot the $10\,\uha$ SE target. More sampled iterations reduce the reported SE but take more time. The CPU implementation of PT2 in TrimCI is also tested; however, the published software reports no SE, and individual repetitions show a significantly larger variance than the targeted SE of GPU implementation and Arrow. Thus, its results are listed separately in Appendix~\ref{app:cpu-configurations} and not included for equal-SE speedup comparison.

\subsection{Parallel performance, memory, and communication}

\begin{figure}[tbp]
 \centering\includegraphics[width=\linewidth]{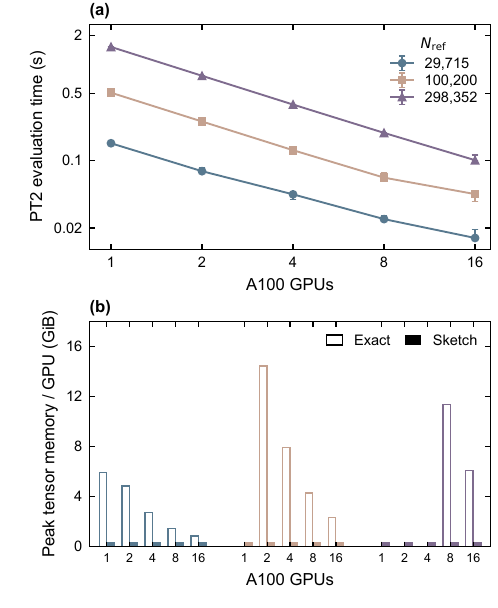}
 \caption{Strong scaling and memory use for N$_2$ references containing approximately 30,000, 100,000, and 300,000 determinants. The benchmark uses a deterministic variant that traverses all sources at a fixed screening threshold, without source sampling or statistical stopping. (a) Evaluation time with sketched accumulation. Symbols show medians of five timing repeats, and error bars span the minimum and maximum. (b) Peak allocated tensor memory per GPU for exact accumulation (open bars), which explicitly sums screened external contributions by determinant, and sketched accumulation (filled bars). Both modes use the same timing scope, defined in Appendix~\ref{app:scaling-protocol}. Colors identify the reference sizes in both panels; missing exact bars indicate unavailable measurements. One to eight A100 GPUs occupy one node, and 16 GPUs occupy two nodes.}
 \label{fig:scaling}
\end{figure}
We assess parallel performance and the benefits of sketch accumulation using a deterministic variant of our method that traverses all reference determinants at a fixed screening threshold. This fixes the workload by removing source sampling and statistical stopping. The sketch mode averages $R=8$ sketch estimates, whereas the exact mode explicitly groups and sums external contributions by determinant. Both modes use the same inputs and screening threshold. The tests use three all-electron N$_2$/cc-pVDZ references (14e28o), containing approximately 30,000, 100,000, and 300,000 determinants, with the published coordinate descent full configuration interaction (CDFCI) Hamiltonian~\cite{cdfci1}. The reported sketch SEs are approximately $30\,\uha$ (Appendix~\ref{app:scaling-protocol}).

Figure~\ref{fig:scaling}a shows the strong scaling. One to eight GPUs occupy one node, and sixteen GPUs occupy two nodes. For the largest reference, the median sketch evaluation time decreases from $1.515$~s on one GPU to $0.194$~s on eight GPUs and $0.101$~s on sixteen GPUs. These correspond to speedups of $7.79\times$ and $14.96\times$, or parallel efficiencies of $97.4\%$ and $93.5\%$. The sixteen-GPU speedups for the two smaller references are $9.61\times$ and $11.27\times$, respectively. The lower efficiencies for smaller references are consistent with collective and kernel-launch overhead accounting for a larger fraction of the evaluation time as excitation work per GPU decreases.

The memory requirements follow from the stored data. In the exact mode, each external record contains a 64-bit determinant key and a complex FP64 partial amplitude, requiring 24 bytes before sorting and communication buffers. The largest reference reaches $3.59\times10^8$ distinct external determinants, so one record per determinant already occupies $8.02$~GiB across all GPUs. In the sketch mode, each bucket stores two complex FP64 values, one per denominator sign. The eight sketches of 65,536 buckets used here therefore occupy $16$~MiB per GPU, independently of the number of external determinants.

For communication, the exact mode first merges repeated keys on each GPU and hashes each remaining key to assign it to a destination GPU. Each GPU sends its key--amplitude records to their assigned destinations, where partial amplitudes for the same determinant are summed across GPUs. On eight GPUs, uniform hash assignment leaves about one eighth of each GPU's records local and sends about one eighth to each of the other seven GPUs. Taking the global distinct-determinant count as a conservative estimate of the total number of locally merged records gives an aggregate sent payload of approximately $7.02$~GiB for the largest reference on eight GPUs. Sketch accumulation instead uses an AllReduce of the $16$-MiB array, corresponding to $224$~MiB sent in aggregate in a ring AllReduce model. The sketch payload is therefore smaller by a factor of approximately 32 in this estimate. Appendix~\ref{app:scaling-protocol} gives the per-GPU and per-destination formulas, including the reference-space reduction retained in both modes.

\begin{figure}[tbp]
\centering\includegraphics[width=\linewidth]{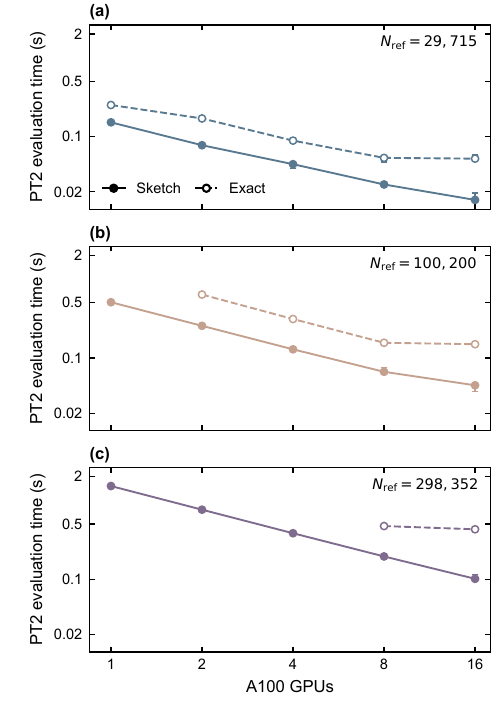}
\caption{Evaluation times using exact (open symbols, dashed lines) and sketched (filled symbols, solid lines) accumulation for N$_2$ references containing (a) 29,715, (b) 100,200, and (c) 298,352 determinants. Both modes use the timing scope defined above. Symbols and ranges are the median and minimum--maximum of five timing repeats. One to eight A100 GPUs occupy one node; 16 GPUs occupy two nodes. Only measured exact results are shown.}
\label{fig:si-sketch-exact}
\end{figure}

Figure~\ref{fig:scaling}b agrees with the memory analysis: exact-mode memory increases with reference size and decreases as the external records are distributed over more GPUs, whereas sketch-mode peak allocated tensor memory remains between $0.308$ and $0.325$~GiB per GPU. These peaks include the reference, integral tables, and working buffers as well as the external accumulators. For the largest reference on eight GPUs, exact accumulation requires $11.35$~GiB per GPU, compared with $0.325$~GiB for sketch accumulation, a reduction by a factor of approximately 35. For all available exact/sketch pairs at the same reference size and GPU count, sketch evaluation is also $1.66$--$4.24$ times faster (Appendix Figure~\ref{fig:si-sketch-exact}).

\begin{figure*}[t]
 \centering\includegraphics[width=\linewidth]{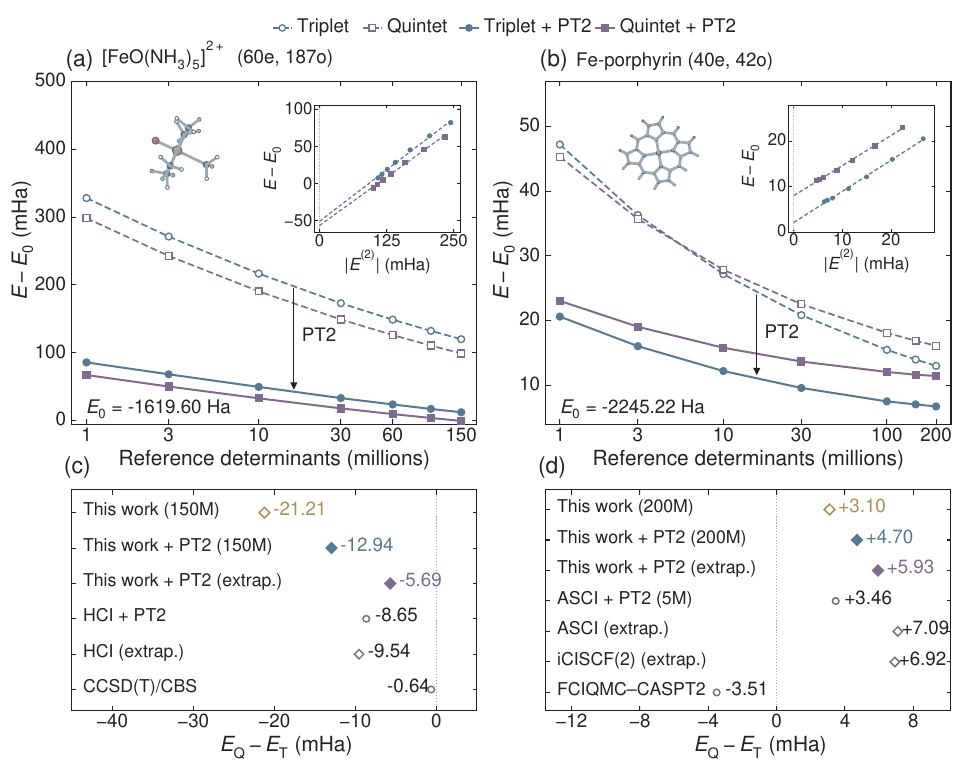}
 \caption{Application to iron complexes. (a,b) Triplet (blue circles) and quintet (purple squares) variational energies (``This work''; open symbols, dashed lines) and PT2-corrected energies (``This work + PT2''; filled symbols, solid lines) for $[\mathrm{FeO}(\mathrm{NH}_3)_5]^{2+}$ in 60e187o and Fe-porphyrin in 40e42o, respectively. Each panel and its inset share the displayed energy offset $E_0$. Dashed lines in the insets show linear extrapolations of SCI+PT2 energies against $|E^{(2)}|$ to zero correction, using all seven reference sizes per state for each complex. (c,d) Gaps $E_{\mathrm Q}-E_{\mathrm T}$ from the largest reference spaces before and after PT2, all-point extrapolation, and representative literature calculations~\cite{DangThesis2023,DrabikSpinCBS2024,LevineASCISCF2020,GuoiCISCF2021,SafariFCIQMCCASPT22025}. Positive gaps favor the triplet. Iron--oxo HCI uses 22e168o and resolution-of-the-identity (RI) integrals; CCSD(T)/CBS includes an outer-core correction. Porphyrin iCISCF(2) uses independently optimized orbitals; FCIQMC--CASPT2 uses a 26e27o space and a different triplet root. The literature points retain these distinct model definitions.}
 \label{fig:science}
\end{figure*}
\begin{figure}[!tbp]
 \centering\includegraphics[width=\linewidth]{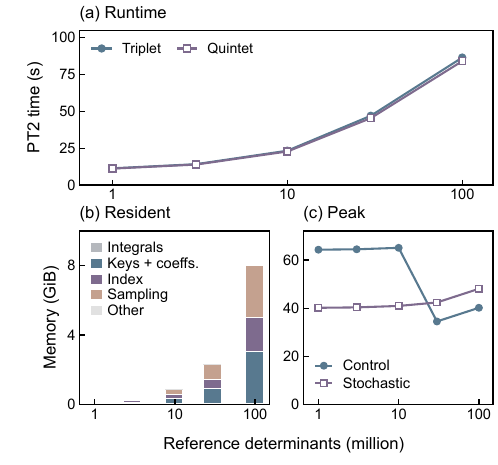}
 \caption{PT2 cost and GPU memory for Fe-porphyrin references containing 1--100 million determinants. (a) Complete PT2 time for the triplet and quintet on one A100-SXM4-80GB. (b) Allocated tensor memory before PT2, decomposed by array sizes. (c) Peak tensor allocation during the deterministic control and stochastic stages. Panels b and c show the triplet; quintet allocations agree at the plotted precision. The control bucket count $B_c$ is selected from the available memory and automatically switched from $2^{31}$ to $2^{30}$ at 30 million determinants, causing the drop in peak memory; the stochastic bucket count remains $2^{28}$. Memory excludes allocator reserve and the CUDA context.}
 \label{fig:fep-cost-memory}
\end{figure}

\subsection{Application to large active space calculations}
\label{sec:large-active-spaces}
We examine two iron complexes whose triplet--quintet splittings test the balance of metal and ligand correlation. Both complexes use PT2 screening thresholds $\epsilon_f=10^{-7}\,\ha$ and $\epsilon_c=10^{-6}\,\ha$. For each state, active--active orbital optimization used TrimCI-COO up to one million reference determinants; subsequent expansion used TrimCI with GPU Hamiltonian application while keeping the orbitals fixed. Throughout this subsection, $\Delta E=E_{\mathrm Q}-E_{\mathrm T}$, so positive values favor the triplet. For extrapolation, each state's SCI+PT2 energies are fitted linearly against $|E^{(2)}|$, and the gap is the difference between the zero-correction intercepts. Stopping criteria and extrapolation procedures are given in Appendix~\ref{app:scientific-protocol}.

The iron(IV)--oxo complex $[\mathrm{FeO}(\mathrm{NH}_3)_5]^{2+}$ is a model of competing triplet and quintet states at a metal--oxo unit. Our calculation uses cc-pVTZ on Fe/O, cc-pVDZ on N, and 6-31G on H. Earlier heat-bath configuration interaction (HCI) calculations froze the N--H $\sigma$ bonds and Fe $3s/3p$ electrons in a 22e168o space, and suggested that correlating the N--H bonds might narrow or reverse the quintet preference~\cite{DangThesis2023}. We include these electrons in a nonrelativistic 60e187o space with conventional four-center integrals, evaluate each state at its published geometry, and expand the references to 150 million determinants, beyond the earlier HCI expansions of 18.9 and 23.9 million. At 150 million determinants, PT2 changes $\Delta E$ from $-21.21$ to $-12.94\,\mha$, and extrapolation using all seven references per state gives $-5.69\,\mha$, compared with the HCI value of $-9.54\,\mha$ (Figure~\ref{fig:science}a,c). The larger calculation keeps the quintet below the triplet and does not produce the suggested reversal.

Neutral iron(II)--porphyrin, FeC$_{20}$H$_{12}$N$_4$, is a minimal heme model whose triplet--quintet ordering is sensitive to the balance of metal and ligand correlation. Using the nonrelativistic 40e42o model of the ASCI-SCF study~\cite{LevineASCISCF2020} in the cc-pVDZ basis, we expand the references to 200 million determinants per state, compared with the reported 5 million. The variational calculations favor the quintet at 1 and 3 million determinants and the triplet from 10 million onward, whereas PT2 favors the triplet at every size (Figure~\ref{fig:science}b). At 200 million determinants, PT2 increases $\Delta E$ from $3.10$ to $4.70\,\mha$. Separate fits to all seven SCI+PT2 energies of each state give an extrapolated gap of $5.93\,\mha$, comparable to the ASCI+PT2 and iCISCF(2) values of $7.09$ and $6.92\,\mha$~\cite{LevineASCISCF2020,GuoiCISCF2021} (Figure~\ref{fig:science}d). These results support the previously reported triplet preference in this model. The corrected gap still changes by $0.11\,\mha$ between 150 and 200 million determinants, so the extrapolated value remains an estimate of the splitting. Fixed-support and screening checks are given in Appendix~\ref{app:application-checks}.

Figure~\ref{fig:fep-cost-memory} summarizes the PT2 cost for the 1--100 million-determinant Fe-porphyrin references. On one A100, complete PT2 evaluation takes $11.2$--$11.4$~s per state at one million determinants and $84.0$--$86.5$~s at 100 million. At 100 million determinants, resident data occupy $8.03$~GiB, including the keys, coefficients, membership index, sampling arrays, and integral tables. Peak tensor allocation reaches $48.04$~GiB with the sketch arrays and temporary workspaces; the sketch arrays are allocated to the largest size the device can hold, and at 30 million determinants the bucket count switches down automatically. The requested SE is $50\,\uha$ at 1M/3M and $25\,\uha$ at larger sizes. All ten calculations meet their respective SE targets after the minimum of six sampling iterations.

\subsection{Application to neural-network reference wave functions}

\begin{figure*}[t]
 \centering\includegraphics[width=\linewidth]{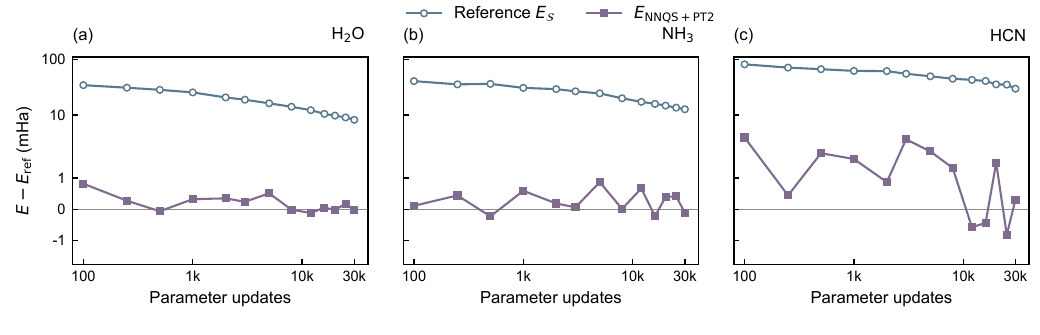}
 \caption{Perturbative improvement of neural-network reference energies for (a) H$_2$O, (b) NH$_3$, and (c) HCN. Open blue circles show the energy $E_{\Sref}$ on the sampled determinant set; filled purple squares show $E_{\mathrm{NNQS+PT2}}$, including the external, raw internal, and Olsen terms. Each pair uses identical determinants and network coefficients. Each panel contains 13 checkpoints from one training trajectory, plotted against the number of parameter updates. Energies are relative to the literature benchmark $E_{\mathrm{ref}}$ defined in Appendix~\ref{app:nqs-reference}. The vertical axis is symmetric logarithmic, with a linear region between $-1.6$ and $+1.6\,\mha$.}
 \label{fig:nqs-checkpoints}
\end{figure*}

The implementation also accepts reference coefficients obtained without CI diagonalization. An autoregressive NNQS~\cite{Wu2023_SC,ShangQiankunNet2025} supplies explicit coefficients on a sampled determinant set, but these coefficients generally leave a nonzero residual within the retained space. Following perturbative residual corrections for variational wave functions~\cite{SharmaStochasticPT2018}, we supplement the external EN-PT2 correction with an internal diagonal-residual term. The combination is nontrivial, and a close inspection produces an Olsen orthogonality correction~\cite{OlsenFCI1990}. Appendix~\ref{app:nqs-reference} gives the expressions and their derivation.

We used QiankunNet wave functions for H$_2$O, NH$_3$, and HCN in cc-pVDZ basis set: all-electron 10e24o for H$_2$O, 8e28o with the N $1s$ core frozen for NH$_3$, and 10e31o with the C/N $1s$ cores frozen for HCN. Water uses the CDFCI Hamiltonian and variational benchmark~\cite{cdfci1}; NH$_3$ and HCN use G2 geometries and SHCI-extrapolated benchmark energies~\cite{YaoSHCI2020}. Each network was trained for 30,000 updates with a nominal support of 30,000 determinants per update. Figure~\ref{fig:nqs-checkpoints} compares the reference and PT2-corrected energies on the same evaluation supports at all 13 recorded checkpoints for each system. At the final checkpoint, the correction reduces the signed errors from $8.23$, $12.69$, and $29.61\,\mha$ to $-0.028$, $-0.114$, and $+0.306\,\mha$, respectively. The PT2-corrected energies reach chemical accuracy ($1.6\,\mha$) while the variational errors are still far outside it, and they remain within that threshold across a range of checkpoints. These results suggest that perturbative post-processing of NNQS~\cite{LiCr2NQS2024,CheDeterministicNNQS2026} may help reduce the variational training needed to reach chemical accuracy.

\section{Discussion and conclusion}
We have implemented fully GPU-accelerated distributed semistochastic EN-PT2 with limited memory use and bounded-size collective communication. Excitation generation, screening, and PT2 weighting are fused into a single GPU traversal. A fixed-size CountSketch avoids explicitly storing the external determinant set. An AllReduce of the sketch arrays followed by a bias-corrected contraction estimates the screened, denominator-regularized correction $F(\epsilon_f)$. The method reaches 93.5\% parallel efficiency in the fixed-workload benchmark, and sketched accumulation reduces peak allocated tensor memory by a factor of approximately 35. At the same requested statistical standard error, the method is 8.21--37.53 times faster than the tested CPU PT2 routine. It is applied to GPU-based SCI methods as well as NNQS reference states.

Currently, the reference state is copied onto each GPU so that membership tests remain local. An 80 GiB device can hold a reference on the order of $10^9$ determinants, depending on the size of the other resident data. When a reference exceeds that capacity, it has to be partitioned across GPUs, changing the membership tests: a GPU that does not hold a determinant must query the GPU that does. In this case, hash-based ownership, Bloom filters for determinants that are certainly absent, and batched queries are ways to limit the membership traffic.

The large reference states used in this work are generated using a hybrid CPU--GPU implementation. With PT2 evaluation and Hamiltonian application running on GPUs, CPU-based excitation generation and determinant selection become the bottleneck. The next step is to move these operations in the variational stage to GPUs.

\section*{Data and software availability}
The numerical source data underlying the figures and tables, together with analysis parameters and figure-reproduction scripts, are available at \url{https://github.com/fulz024/2026-gpu-pt2-data-deposit}. The software implementation is hosted at \url{https://github.com/fulz024/QiankunPT}. Arrow and TrimCI are available from their cited repositories~\cite{ArrowSoftware2026,TrimCISoftware2026}. The additional Figure~\ref{fig:science} data for this preprint are supplied as ancillary files.

\section*{Acknowledgments}
The authors acknowledge support from the National Natural Science Foundation of China (Grant No. 92570112).

\Needspace{7\baselineskip}
\paragraph{AI usage disclosure.}
OpenAI Codex and Cursor assisted with software inspection, analysis, literature-search organization, language editing, and figure preparation. The authors are responsible for checking the scientific content, numerical evidence, citations, and final manuscript.

\FloatBarrier
\qkbibliography{manuscript,stochastic_references}

\clearpage
\nobalance
\qkappendix
\iflatexml
\section*{Appendices}
\else
\begin{qkfrontmatter}
\pdfbookmark[0]{Appendices}{appendices}
{\centering\fontsize{16}{19}\selectfont\bfseries\color{QKPrimary}Appendices\par}
\end{qkfrontmatter}
\thispagestyle{qkbody}
\fi
\numberwithin{equation}{section}
\numberwithin{figure}{section}
\numberwithin{table}{section}
\section{Estimator and statistical error}
\label{app:estimator}
The deterministic control and sampled correction follow semistochastic heat-bath configuration interaction (SHCI)~\cite{shci,LiFastSHCI2018}. The fully GPU-accelerated implementation evaluates their excitation contributions through fused generation and direct accumulation in random-phase CountSketch arrays (Sections~\ref{sec:coupled-kernel} and~\ref{sec:sketch}).

Screening uses the original $H_{ai}c_i$, before source reweighting. The single-sketch calculations use the sign-preserving denominator floor
\begin{equation}
 \begin{gathered}
\widetilde d_a=
 \begin{cases}
 d_a, & |d_a|\geq\eta,\\
 +\eta, & 0\leq d_a<\eta,\\
 -\eta, & -\eta<d_a<0,
 \end{cases}
 \\
  \eta=10^{-10}\,\ha.
 \end{gathered}
\end{equation}

\subsection{Source sampling and sketch accumulation}
\label{app:sampled-reference}
\label{app:bucket-arrays}
Using the fixed subset $\Dref$, sampled complement $\Tref$, and weights $w_i$ defined in Section~\ref{sec:method}, set $x_{ai}=t_{ai}^{\epsilon}/\sqrt{|\widetilde d_a|}$ and $Q(v)=\sum_a\operatorname{sgn}(\widetilde d_a)|v_a|^2$. Let $y=\sum_{i\in\Dref}x_i$, $z=\sum_{i\in\Tref}w_i x_i$, and $S=\sum_{i\in\Tref:n_i>0}\gamma_iw_i^2Q(x_i)$. The bias-corrected source estimator is
\begin{equation}
 \begin{gathered}
\widehat F_{\rm src}(\epsilon)=\frac{1}{\normc}
 \left[Q(y+z)+\frac{Q(z)}{m-1}-S\right],\\
 \gamma_i=1-\frac{mp_i}{n_i}+\frac{1}{m-1}\quad(n_i>0).
 \label{eq:source-estimator}
 \end{gathered}
\end{equation}
For $X=\sum_{i\in\Tref}x_i$ and $J=\sum_{i\in\Tref}Q(x_i)/p_i$, the multinomial sampling moments give
\begin{equation}
 \begin{gathered}
\mathbb E[z]=X,\\
 \mathbb E[Q(z)]=\frac{m-1}{m}Q(X)+\frac{J}{m},
 \\
  \mathbb E[S]=\frac{J}{m-1}.
 \end{gathered}
\end{equation}
Substitution yields $\mathbb E[\widehat F_{\rm src}]=Q(y+X)/\normc=F(\epsilon)$, as in the SHCI estimator~\cite{shci,LiFastSHCI2018}.

For threshold $\ell\in\{f,c\}$, restrict these sums to the selected external batch and form sketches $A_{\ell,D}=\mathcal K y_\ell$ and $A_{\ell,T}=\mathcal K z_\ell$. With the sketch contraction $Q(A)$ of Eq.~\eqref{eq:sketch-expectation},
\begin{equation}
 \begin{aligned}
 \widehat F_\ell&=\frac{P}{\normc}\Bigl[Q(A_{\ell,D}+A_{\ell,T})\\
 &\qquad+\frac{Q(A_{\ell,T})}{m-1}-S_\ell\Bigr],\\
 \Delta_w&=\widehat F_f-\widehat F_c.
 \end{aligned}
 \label{eq:kernel-energy}
\end{equation}
The self term $S_\ell$ is accumulated directly, without sketching. Fine and coarse estimates share the source draw, external batch, bucket map, and phase map. Under the ideal random-map assumptions of Eq.~\eqref{eq:sketch-expectation}, $\mathbb E[\Delta_w]=F(\epsilon_f)-F(\epsilon_c)$. The deterministic control instead uses all sources with unit weights and $P=1$.

\subsection{Reported standard error}
\label{app:variance}
For the production setting $R=1$, let $C_k$ be independent deterministic-control estimates and $\Delta_w$ independent sampled corrections. For $n_c$ control evaluations and $W$ sampling iterations,
\begin{equation}
 \begin{gathered}
\widehat F=\overline C+\overline\Delta,\\
 \mathrm{SE}^2=\max\!\left(\frac{s_C^2}{n_c},V_c^{\rm guard}\right)+\frac{s_\Delta^2}{W}.
 \label{eq:se}
 \end{gathered}
\end{equation}
For $n_c=1$, the empirical control-variance term is zero, but the norm-based bound $V_c^{\rm guard}$ remains. Control maps are independent of the sampled correction; each sampling iteration refreshes its source draw, external batch, and sketch.

For control amplitudes $\alpha_a=\sum_i t_{ai}^{\epsilon_c}/\sqrt{|\widetilde d_a|}$ and $B_c$ buckets per denominator sign, define $A=\normc^{-1}\sum_a|\alpha_a|^2$ and $Z_k=\normc^{-1}\sum_{s,b}|(\mathcal K_k\alpha)_{s,b}|^2$. Ideal uniform bucket maps and independent fourth-root phases give $\operatorname{Var}(Z_k)\leq A^2/B_c$, also bounding the signed energy variance. Chebyshev's inequality gives
\begin{equation}
 \begin{gathered}
A\leq A_n^+=\frac{\overline Z_n}{1-r_n},\\
 r_n=(B_c n\delta)^{-1/2}<1,\\
 V_c^{\rm guard}=\frac{(A_n^+)^2}{B_c n}.
 \label{eq:norm-guard}
 \end{gathered}
\end{equation}
At fixed $n$, the bound fails with probability at most $\delta$. The implementation uses $\delta=10^{-3}/16$ and at most 16 evaluations; a union bound applies to these possible stopping points under the same random-map assumptions.

\subsection{Empirical variance checks}
\label{app:empirical-variance}
\begin{figure}[tbp]
 \centering\includegraphics[width=\linewidth]{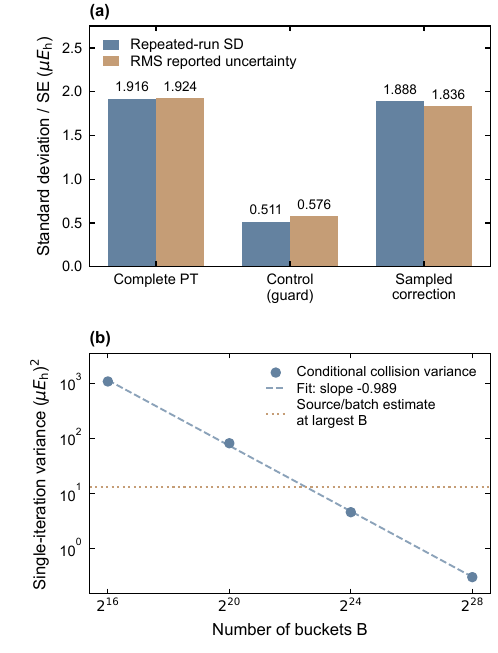}
 \caption{Statistical-error checks for the Fe-porphyrin triplet with a fixed one-million-determinant reference. (a) Empirical standard deviations from 64 independent complete calculations compared with root-mean-square reported uncertainties. The control entry uses the conservative variance guard; the sampled-correction entry refers to the mean of six iterations. (b) Conditional collision variance averaged over eight source/batch combinations, each evaluated with sixteen hash maps. The dashed line is the fitted power law; the dotted line marks the source/batch variance estimate at the largest $B$.}
 \label{fig:si-variance-validation}
\end{figure}

We use the fixed (40e, 42o) Fe-porphyrin triplet Hamiltonian and its one-million-determinant reference. The 64 complete calculations refresh all random streams independently and use $R=1$, $P=16$, $U=262144$, $K=65536$, $m=233501$, $B_c=2^{31}$, and $B=2^{28}$, with screening thresholds $\epsilon_f=10^{-7}\,\ha$ and $\epsilon_c=10^{-6}\,\ha$. Each calculation requests an SE of $50\,\uha$ and stops after the minimum six sampled-correction iterations. The empirical standard deviation of the complete correction is $1.9163\,\uha$, compared with a root-mean-square reported SE of $1.9239\,\uha$. For the deterministic control, the empirical standard deviation and root-mean-square guard are $0.5110$ and $0.5763\,\uha$, respectively; the corresponding values for the averaged sampled correction are $1.8882$ and $1.8355\,\uha$ (Figure~\ref{fig:si-variance-validation}a).

To separate collision variance from source-sampling and external-batch variance, we evaluate eight independent source/batch combinations with the same sixteen hash seeds at each $B\in\{2^{16},2^{20},2^{24},2^{28}\}$, giving 512 sampled-correction estimates. Fine and coarse thresholds share the source draw, external batch, and hash maps, so the measured quantity is their paired difference, $\Delta_{dh}$, for source/batch combination $d$ and hash seed $h$. Let $n_d=8$, $n_h=16$, and let $\overline\Delta_{d\cdot}$, $\overline\Delta_{\cdot h}$, and $\overline\Delta$ denote the row, column, and overall means. Because the hash seeds are shared across source/batch combinations, we use the crossed-design estimates
\begin{align}
 \widehat V_{\rm hash}&=\frac1{n_d}\sum_d\operatorname{Var}_h(\Delta_{dh}),\\
 M_{\rm int}&=\frac{\sum_{d,h}(\Delta_{dh}-\overline\Delta_{d\cdot}-\overline\Delta_{\cdot h}+\overline\Delta)^2}{(n_d-1)(n_h-1)},\\
 \widehat V_{\rm sample}&=\operatorname{Var}_d(\overline\Delta_{d\cdot})-\frac{M_{\rm int}}{n_h},
 \label{eq:empirical-variance-components}
\end{align}
where each $\operatorname{Var}$ is an unbiased sample variance. At $B=2^{28}$, these estimates are $\widehat V_{\rm hash}=3.008\times10^{-13}\,\ha^2$ and $\widehat V_{\rm sample}=1.289\times10^{-11}\,\ha^2$: collisions contribute approximately 2.3\% of their sum. The fitted slope of $\log\widehat V_{\rm hash}$ against $\log B$ is $-0.9893$ (Figure~\ref{fig:si-variance-validation}b). This decomposition explains the measured sampled-correction variance; the observed iteration variance in Eq.~\eqref{eq:se} already includes both contributions.

\section{Benchmark settings and convergence checks}
\label{app:protocol}
\subsection{Sampling parameters and timing}
\label{app:production-policy}
\label{sec:production-policy}
The four 2M-reference performance calculations use $R=1$, $B_c=2^{31}$, $B=2^{28}$, $U=262144$, $K=65536$, $P=16$, and $\epsilon_c=10\epsilon_f$. Here $U$ is the expected number of distinct sources, including the $K$ fixed sources. The draw count is the smallest $m\geq2$ satisfying
\begin{equation}
 K+\sum_{i\in\Tref}[1-(1-p_i)^m]\geq U.
 \label{eq:occupancy}
\end{equation}
Table~\ref{tab:references} lists the resulting draw and iteration counts. The sampled correction uses at least six iterations and stops at a total reported SE of $10\,\uha$. Control evaluations continue until their variance bound is below half the requested variance budget, up to 16 evaluations; all four runs required one.
\begin{table}[!htbp]
\centering\small
\caption{GPU sampling parameters for the 2M-reference performance calculations. $m$ is the number of draws per iteration and $W$ the number of sampled-correction iterations.}
\label{tab:references}
\begin{tabular}{lrrr}\toprule
System & $\epsilon_f$ ($\ha$) & $m$ & $W$\\\midrule
Cr$_2$ & $5\times10^{-8}$ & 212316 & 6\\
Benzene & $5\times10^{-8}$ & 211755 & 6\\
Pyrazine & $1\times10^{-7}$ & 219482 & 7\\
Pyrimidine & $1\times10^{-7}$ & 221348 & 46\\
\bottomrule\end{tabular}
\end{table}

Complete PT2 timing includes sampling setup, the deterministic control, sampled correction, communication, and energy evaluation, with GPU synchronization at both boundaries. Input parsing, integral-table construction, index preparation, and upload are excluded. The GPU and CPU performance runs have 48- and 900-GiB host allocations, respectively; Arrow uses an 800-GB program budget. Additional validation uses 32 Arrow cores with 32 GiB for small systems and 128 GiB for N$_2$, allocating 80\% to Arrow. The Cr$_2$ and validation N$_2$ bond lengths are $1.68$ and $1.10$~\AA, respectively; other molecular definitions are given in the main text.

\subsection{CPU comparison}
\label{app:cpu-configurations}
Arrow's fast SHCI implementation (commit \texttt{0174c4a}) uses its three-stage PT2 estimator, original stopping rule, and PT2 memory fraction 0.4. Trimmed configuration interaction (TrimCI; commit \texttt{0deaee9}) uses its original \texttt{compute\_pt2} routine: sketch width 262144, depth 5, norm target 0.99, stochastic fraction 0.02, Bloom setting 0.001, and intruder threshold $10^{-6}\,\ha$.

TrimCI screens same-spin connections using separate Coulomb/exchange magnitudes, whereas Arrow and the GPU implementation screen the matrix-element magnitude, so equal nominal thresholds retain different connections. TrimCI returns no total SE. Four independent 16-core benzene runs give a single-run sample SD of $399.18\,\uha$.

\begin{table*}[!tp]
\centering\small
\caption{Complete PT2 evaluation times and reported statistical SEs. All references contain 2M determinants. The GPU implementation and Arrow use the same requested SE of $10\,\uha$. TrimCI is included as an additional CPU timing reference; it reports no total SE.}
\label{tab:performance}
\setlength{\tabcolsep}{4pt}
\begin{tabular}{lrrrrrr}\toprule
 & \multicolumn{3}{c}{PT2 time (s)} & \multicolumn{2}{c}{Reported SE ($\uha$)} & Arrow/\\
System & This work & Arrow & TrimCI & This work & Arrow & This work\\\midrule
Cr$_2$ & 37.70 & 733.18 & 294.96 & 4.10 & 4.93 & 19.45\\
Benzene & 79.01 & 648.52 & 740.19 & 2.98 & 8.00 & 8.21\\
Pyrazine & 77.81 & 2920.73 & 598.95 & 9.46 & 9.98 & 37.53\\
Pyrimidine & 727.58 & 12318.14 & 1479.11 & 9.95 & 9.99 & 16.93\\
\bottomrule\end{tabular}\end{table*}

\subsection{Fixed-workload scaling measurements}
\label{app:scaling-protocol}
The scaling measurements use a deterministic variant of the GPU implementation that traverses all reference determinants at a fixed screening threshold, without source sampling or statistical stopping. The system is all-electron N$_2$/cc-pVDZ (14e28o), with the published coordinate descent full configuration interaction (CDFCI) Hamiltonian at a bond length of $2.118\,a_0$~\cite{cdfci1}. The tests use $R=8$ sketches, $B=65536$ buckets per denominator sign, and source batches of 512. The matrix-element cutoff and denominator floor are both $10^{-8}\,\ha$. Both accumulation modes use identical source sets and cutoffs; sketch seeds are fixed across timing repeats. The timed evaluation includes external excitation generation and accumulation, the reference-space Hamiltonian action $H_{\Sref\Sref}c$, and the projected internal residual correction, including their collectives. Once $H_{\Sref\Sref}c$ is available, the residual and its diagonal correction are evaluated by vector operations. Setup, sampling, compilation, upload, and two warm-up calls are excluded. Each repeat takes the maximum time over ranks. Memory is peak allocated tensor memory per GPU, excluding the CUDA context and other untracked allocations.

The reported SE is the sample standard deviation of the eight sketch estimates divided by $\sqrt{8}$. All sources are enumerated, so it measures sketch variation without source-sampling noise. Table~\ref{tab:scaling-SE} lists the SE and the observed difference from exact accumulation at the same screening threshold. Results on 1, 2, 4, 8, and 16 GPUs agree at the precision shown. The five timing repeats reuse the same hashes and are not additional independent estimates for reducing the SE.
\begin{table*}[!tp]
\centering\small
\caption{Reported sketch SEs and energy differences in the fixed-workload calculations. Differences are absolute deviations from the external correction obtained by explicit accumulation at the same screening threshold. Values are from eight GPUs; all tested GPU counts give the same rounded values.}
\label{tab:scaling-SE}
\begin{tabular}{rrr}\toprule
$N_{\rm ref}$ & Reported sketch SE ($\uha$) & Difference from exact ($\uha$)\\\midrule
29,715 & 27.78 & 24.90\\
100,200 & 29.63 & 17.21\\
298,352 & 31.74 & 9.61\\
\bottomrule\end{tabular}
\end{table*}

In these single-word determinant tests, an explicit external record consists of one 8-byte key and a 16-byte complex amplitude. Locally repeated keys are combined before the exact-mode all-to-all exchange, but a determinant reached on several GPUs still contributes several partial records. For the largest reference, the globally combined external set contains 358,999,436 determinants, corresponding to $8.024$~GiB for one key--amplitude record per determinant before auxiliary buffers. With $R$ sketches, $B$ buckets per sign, two denominator signs, and complex FP64 storage, the external sketch allocation is
\begin{equation}
 C_{\rm sketch}=R\times2\times B\times16=32RB\quad\text{bytes per GPU}.
\end{equation}
Thus $R=8$ and $B=65536$ give $16$~MiB per GPU, or $128$~MiB across eight GPUs. These fixed-threshold tests use one amplitude array per sketch, rather than the four arrays of the sampled correction in Section~\ref{sec:cost-scheduler}. The external array size is independent of the number of generated connections or distinct external determinants. Reference storage, integral tables, and batch workspaces remain additional memory costs.

In the exact mode, GPU $g$ first combines repeated local keys and retains $L_g$ key--amplitude records. A hash of each key assigns a destination GPU $k$. GPU $g$ sends one record for each locally retained key assigned to $k$, and GPU $k$ adds the partial amplitudes for the same key received from different GPUs. Let $n_{gk}$ be the number of records sent from $g$ to $k$, with $L_g=\sum_k n_{gk}$. Records with $k=g$ stay local. Each GPU first exchanges an 8-byte record count with each peer, followed by its packed key--amplitude records. Thus the payload from $g$ to another GPU $k$ is $8+24n_{gk}$ bytes, and the aggregate sent payload is
\begin{equation}
 V_{\rm exact}^{\rm sent}=8G(G-1)+24\sum_{g\ne k}n_{gk}
 \quad\text{bytes}.
\end{equation}
Here $G$ is the number of GPUs; protocol overhead is excluded. Under approximately uniform hash assignment, $n_{gk}\simeq L_g/G$, giving
\begin{equation}
 V_{\rm exact}^{\rm sent}\simeq8G(G-1)+24\frac{G-1}{G}\sum_g L_g
 \quad\text{bytes}.
\end{equation}
The global distinct-determinant count $U_{\rm ext}$ satisfies $U_{\rm ext}\leq\sum_g L_g$, because a key reached on several GPUs contributes one local record on each. Taking $\sum_g L_g=U_{\rm ext}=358999436$ as a conservative record-count example gives approximately $7.02$~GiB sent in aggregate on eight GPUs. With balanced local record counts, this example corresponds to about $128.39$~MiB from each GPU to each of its seven peers, or $898.72$~MiB sent per GPU. These are uniform-hash estimates: the original reports retain $U_{\rm ext}$ but not the per-destination counts $n_{gk}$.

For the sketch mode, a ring AllReduce model~\cite{NCCLTestsPerformance} gives
\begin{equation}
 V_{\rm sketch}^{\rm sent}=\frac{2(G-1)}{G}C_{\rm sketch}
 \quad\text{bytes per GPU},
\end{equation}
or $28$ and $30$~MiB per GPU on eight and sixteen GPUs, respectively. The aggregate sent payload is $2(G-1)C_{\rm sketch}$, or $224$~MiB on eight GPUs. Compared with the conservative exact-mode record-count example above, this is approximately 32 times smaller. These expressions count data payloads rather than measured link traffic. Both benchmark modes also reduce the complex reference-space vector $H_{\Sref\Sref}c$, whose input array occupies $16N_{\rm ref}$ bytes ($4.55$~MiB for the largest reference), and scalar counters. Sketching removes the external determinant-record exchange while retaining these common reductions.

\subsection{SCI convergence and extrapolation}
\label{app:scientific-protocol}
\label{app:application-checks}
The iron--oxo basis is cc-pVTZ on Fe/O, cc-pVDZ on N, and 6-31G on H; Fe-porphyrin uses cc-pVDZ.

Active--active orbital optimization stops at an effective accepted-step energy change below $10^{-8}\,\ha$ or a gradient norm below $10^{-5}$. All four 1M iron-complex references stopped by the energy criterion, without establishing gradient stationarity; larger references keep these orbitals fixed. Each CI reference satisfies the selected-space residual $\|H_{\Sref\Sref}c-Ec\|_2\leq10^{-7}\,\ha$ for normalized $c$. Requested PT2 SEs are $50\,\uha$ at 1M/3M and $25\,\uha$ at larger sizes.

At 30M Fe-porphyrin determinants, tightening the CI tolerance to $10^{-8}\,\ha$ gives triplet\slash quintet residuals of $7.16\times10^{-9}$/$7.98\times10^{-9}\,\ha$ and changes the variational gap by $1.08\times10^{-10}\,\ha$. Halving the PT2 screening threshold changes the corrected gap from $4.08851$ to $4.06535\,\mha$. These checks use fixed 30M supports and orbitals. The combined reported PT2 SE is $0.03432\,\mha$ for 200M Fe-porphyrin and $0.03534\,\mha$ for the 150M iron--oxo pair.

At fixed orbitals and basis, each state's energies are fitted by unweighted least squares to $E_{\mathrm{SCI+PT2}}=A+B|E^{(2)}|$; the extrapolated gap is $A_{\mathrm Q}-A_{\mathrm T}$. All sizes are used: 1M, 3M, 10M, 30M, 60M, 100M, and 150M for the iron--oxo complex, and 1M, 3M, 10M, 30M, 100M, 150M, and 200M for Fe-porphyrin.

The earlier iron--oxo heat-bath configuration interaction (HCI) results use resolution-of-the-identity (RI) integrals, whereas the present results use conventional integrals~\cite{DangThesis2023}. Porphyrin ASCI+PT2 uses the supplied orbital subspaces before our active--active optimization~\cite{LevineASCISCF2020}; iCISCF(2) optimizes its orbitals independently~\cite{GuoiCISCF2021}. FCIQMC--CASPT2 uses a different 26e27o ANO-RCC model and $^3E_g$ triplet~\cite{SafariFCIQMCCASPT22025}. Our $C_1$ calculations start from the ASCI $^3A_{2g}$ lineage without imposing that spatial symmetry; both SCI and PT2 remain within the stated 40e42o space.

\section{Neural-network reference wave functions}
\label{app:nqs-reference}
\subsection{Internal residual and Olsen correction}
For normalized network coefficients $x_i=c_i/\sqrt{\normc}$ on the retained set,
\begin{equation}
 \begin{gathered}
E_{\Sref}=x^\dagger H_{\Sref\Sref}x,\\
 r_i=(H_{\Sref\Sref}x)_i-E_{\Sref}x_i.
 \label{eq:nqs-internal-residual}
 \end{gathered}
\end{equation}
A projected diagonal partition makes $x$ an eigenvector of the zeroth-order Hamiltonian~\cite{SharmaStochasticPT2018}:
\begin{equation}
 \begin{gathered}
H_0=E_{\Sref}P_x+Q_xH_{\mathrm{diag}}Q_x,\\
 P_x=|x\rangle\langle x|,\quad Q_x=1-P_x.
 \label{eq:nqs-partition}
 \end{gathered}
\end{equation}
With $\widetilde d_i$ obtained by regularizing $E_{\Sref}-H_{ii}$, the raw internal term and two contractions are
\begin{equation}
 \begin{gathered}
 E_{\mathrm{int,raw}}^{(2)}=\sum_{i\in\Sref}\frac{|r_i|^2}{\widetilde d_i},\\
 A=\sum_{i\in\Sref}\frac{x_i^*r_i}{\widetilde d_i},\qquad
 B=\sum_{i\in\Sref}\frac{|x_i|^2}{\widetilde d_i}.
 \end{gathered}
 \label{eq:nqs-raw-internal}
\end{equation}
For $B\ne0$, enforcing orthogonality after diagonal preconditioning gives~\cite{OlsenFCI1990,SleijpenVanDerVorst1996}
\begin{equation}
 \begin{gathered}
t_i=\frac{r_i+\lambda x_i}{\widetilde d_i},\\
 \lambda=-\frac{A}{B},\\
  x^\dagger t=0,
 \label{eq:nqs-olsen-vector}
 \end{gathered}
\end{equation}
and hence
\begin{equation}
 \begin{aligned}
 E_{\mathrm{Olsen}}^{(2)}&=-\frac{|A|^2}{B},\\
 E_{\mathrm{int,proj}}^{(2)}&=r^\dagger t\\
 &=E_{\mathrm{int,raw}}^{(2)}+E_{\mathrm{Olsen}}^{(2)}.
 \end{aligned}
 \label{eq:nqs-olsen-energy}
\end{equation}
The full correction combines the external EN sum with this projected internal term:
\begin{equation}
 \begin{aligned}
 E_{\mathrm{NNQS+PT2}}&=E_{\Sref}+E_{\mathrm{ext}}^{(2)}\\
 &\quad+E_{\mathrm{int,raw}}^{(2)}+E_{\mathrm{Olsen}}^{(2)}.
 \end{aligned}
 \label{eq:nqs-full-correction}
\end{equation}
For a converged CI eigenvector, $r=A=0$, leaving only the external correction. The perturbative energy need not be variational.

\balance
\subsection{Training and energy benchmarks}
The network has four Transformer blocks, width 32, four attention heads, zero dropout, and a phase network with four hidden layers of width 512~\cite{ShangQiankunNet2025}. Configuration interaction singles and doubles (CISD)-supervised initialization precedes 30,000 parameter updates, each using one fresh batched autoregressive sampling (BAS) support of nominal size 30,000. AdamW uses $\beta_1=0.9$, $\beta_2=0.99$, $\epsilon=10^{-9}$, and zero weight decay. The capped Transformer learning-rate schedule has 4000 warm-up steps, scale factor 1, and maximum rate $10^{-4}$.

All 13 recorded checkpoints of each trajectory are evaluated on their own sampled supports, with the frozen network coefficients normalized on that support and no CI diagonalization. External corrections are deterministic screened sums. Table~\ref{tab:nqs-benchmarks} lists the final support sizes and literature benchmarks used to define the plotted errors.
\begin{table}[!htbp]
\centering\small
\setlength{\tabcolsep}{3.5pt}
\caption{Final NNQS evaluation supports and benchmark energies. H$_2$O uses the CDFCI variational benchmark~\cite{cdfci1}; NH$_3$ and HCN use SHCI extrapolations~\cite{YaoSHCI2020}. The dash denotes an unreported uncertainty.}
\label{tab:nqs-benchmarks}
\begin{tabular}{lrrr}\toprule
System & $N_{\rm ref}$ & $E_{\rm ref}$ ($\ha$) & Uncertainty ($\uha$)\\\midrule
H$_2$O & 30,861 & $-76.2418601$ & ---\\
NH$_3$ & 29,199 & $-56.402476876204$ & 5.78\\
HCN & 29,170 & $-93.189884178140$ & 7.19\\
\bottomrule\end{tabular}
\end{table}

\subsection{Component comparison}
Table~\ref{tab:hcn-components} separates the corrections on the final 29,170-determinant HCN support. The retained-space matrix-element cutoff and denominator floor are $10^{-8}\,\ha$. The Olsen term raises the energy by $0.0853575\,\mha$ in this example, slightly increasing the error relative to the raw internal correction.
\begin{table}[!htbp]
\centering\small
\setlength{\tabcolsep}{3.5pt}
\caption{HCN energies from identical network coefficients and support. Errors are relative to the benchmark in Table~\ref{tab:nqs-benchmarks}.}
\label{tab:hcn-components}
\begin{tabularx}{\linewidth}{@{}>{\raggedright\arraybackslash}Xrr@{}}\toprule
Included corrections & Energy ($\ha$) & Error ($\mha$)\\\midrule
Reference & $-93.160276092639$ & 29.60809\\
External & $-93.184458694913$ & 5.42548\\
External + raw internal & $-93.189663815743$ & 0.22036\\
External + raw internal + Olsen & $-93.189578458265$ & 0.30572\\
\bottomrule\end{tabularx}
\end{table}

\FloatBarrier
\end{document}